\documentclass[sigconf]{acmart}

\usepackage{multirow}
\usepackage{subcaption}
\usepackage{tabularx}
\usepackage{xcolor}
\usepackage{fancybox}
\usepackage{fontawesome5}

\newcommand{\summarybox}[2]{%
  \par\medskip\noindent
  \setlength{\fboxrule}{1.3pt}%
  \setlength{\fboxsep}{0pt}%
  \fcolorbox{yellow!70!gray}{yellow!10}{%
    \parbox{\linewidth}{%
      \begingroup
      \setlength{\fboxsep}{0pt}
      \colorbox{yellow!70!gray}{%
        \parbox{\linewidth}{%
          \vspace{4pt}
          \hspace*{7pt}
          \textbf{\color{black}\faLightbulb\ \ #1}%
          \vspace{4pt}
        }%
      }%
      \endgroup
      \par
      \vspace{5pt}%
      \hspace*{7pt}%
      \parbox{\dimexpr\linewidth-14pt\relax}{#2}%
      \vspace{7pt}%
    }%
  }%
  \par\medskip
}

\AtBeginDocument{%
  }

\copyrightyear{2026}
\acmYear{2026}
\setcopyright{cc}
\setcctype{by}
\acmConference[MSR '26]{23rd International Conference on Mining Software Repositories}{April 13--14, 2026}{Rio de Janeiro, Brazil}
\acmBooktitle{23rd International Conference on Mining Software Repositories (MSR '26), April 13--14, 2026, Rio de Janeiro, Brazil}
\acmPrice{}
\acmDOI{10.1145/3793302.3793592}
\acmISBN{979-8-4007-2474-9/2026/04}

\newboolean{showcomments}
\setboolean{showcomments}{true}

\ifthenelse{\boolean{showcomments}}
{\newcommand{\nbc}[3]{
 {\colorbox{#3}{\bfseries\sffamily\scriptsize\textcolor{white}{#1}}}
 {\textcolor{#3}{\sf\small$\blacktriangleright$\textit{#2}$\blacktriangleleft$}}
 }
}
{\newcommand{\nbc}[3]{}
}

\newcommand{\rqi}{Why developers reject AI-agents generated fixes?}
\newcommand{\rqii}{How different are rejection reasons in terms of code changes?}

\begin{document}

\title{Understanding the Rejection of Fixes Generated by Agentic Pull Requests - Insights from the AIDev Dataset}


\author{Mahmoud Abujadallah}
\affiliation{%
  \institution{École de Technologie Supérieure}
  \city{Montréal}
  \country{Canada}
}
\email{mahmoud.abujadallah.1@ens.etsmtl.ca}

\author{Ali Arabat}
\affiliation{%
  \institution{École de Technologie Supérieure}
  \city{Montréal}
  \country{Canada}
}
\email{ali.arabat.1@ens.etsmtl.ca}

\author{Mohammed Sayagh}
\affiliation{%
  \institution{École de Technologie Supérieure}
  \city{Montréal}
  \country{Canada}
}
\email{mohammed.sayagh@etsmtl.ca}

\renewcommand{\shortauthors}{Abujadallah et al.}

\begin{abstract}
AI coding agents are increasingly used to generate pull requests (PRs) that propose code fixes in software projects. From a first exploration of the AIDev dataset, we find that 46.41\% of the fixes proposed by the agents Copilot, Devin, Cursor, and Claude are rejected. This represents a significant amount of wasted resources that require human reviews, verifications, and running tests and validations for fixes that are merely discarded. Our goal in this paper is to understand the failure modes of AI-agents, an understanding that is crucial for better integrating AI-agents as efficient teammates. In this paper, we conduct a qualitative study on a representative sample of 306 non-merged pull requests created or co-authored by the agents mentioned earlier, followed by a quantitative analysis of the reasons for rejection. Our qualitative findings identify 14 reasons divided into four high-level categories for rejecting AI-agent fixes. We observe that developers can reject fixes due to fixes whose implementation is incorrect (e.g., incomplete, wrong approach), fixes that do not pass the continuous integration (CI) pipelines and fail tests, fixes for which the agent is unable to perform the implementation (e.g., no code generated, sessions lost), and fixes whose priority is low. Our results shed light on the importance of better guiding the model at these levels: (1) proposing hints about the approach to follow for fixing an issue, (2) outlining constraints or limitations regarding the approaches that should not be taken, and (3) instructing the agent on how to validate the implementation through CI pipelines and without introducing a breaking change. Our results suggest the need for good prioritization of tasks so that generated fixes do not lead to wasted human review efforts or wasted agent resources (e.g., tokens, compute, or allowed number of requests).

\end{abstract}

\ccsdesc[500]{Software and its engineering~Collaboration in software development}
\keywords{AI-agents, Agentic-PRs, Fixing issues, Qualitative study}

\received{30 December 2025}
\received[revised]{19 January 2026}
\received[accepted]{19 January 2026}

\maketitle


\section{Introduction}

AI-agents are gaining an important place in software development~\cite{li2025aiteammates_se3}. They are part of software teams and can fully participate in the development of new features, fixing bugs, and improving existing code~\cite{li2025aiteammates_se3}
. These tasks arise from natural language prompts in which a developer initiates a task for an AI-agent to complete.

Toward a vision of AI-agents being completely autonomous and capable of participating as productive and efficient contributors, we should achieve a deep understanding of the limitations and the failure modes of AI-agents. From the AIDev dataset, we observe that 46.41\% of the fixes created or co-authored by Copilot, Devin, Cursor, or Claude end up being rejected. Understanding the reasons behind these rejections is critical to better guide practitioners on how to use agents and inform future studies on how to help developers better integrate these AI-agents. Besides, for certain LLMs, the generated code that ends up being rejected wastes AI-agent tokens or limited premium queries. Another motivation for the need to understand the failing modes of AI-agents is the trustworthiness of agents; lack of trust can affect the adoption of AI-agents~\cite{khati2025trustworthiness,balayn2024empirical}. 

In this paper, we study the reasons why developers reject AI-agent fixes and quantify the effort of review caused by each reason, seen as a wasted effort. The effort we measure through the code churn and the number of comments. The higher the code churn, the more effort one has to investigate on the generated code before reaching the conclusion of rejecting the fix. We focus on fixes for several reasons: (1) fixing bugs is about restoring the correct behavior of software; hence, their rejection is more likely to be grounded in technical issues rather than users, developers', or product-based preferences; (2) the tasks are more likely to be scoped around a given fine-grained task of fixing a failure; and (3) AI-agent pull requests are second in terms of dominance, yet first in terms of rejection rate according to the AI-Dev dataset. 

The closest work to our paper is a recent study~\cite{watanabe2025use} that analyzed the rejection reasons for Claude Agentic-PRs across heterogeneous task types. Our work differs fundamentally in three key ways. (1) Their analysis is limited to a single agent (Claude), while we look at four agents. (2) Most importantly, their analysis aggregate pull requests with different intents (e.g., refactoring, new features, documentation, etc.), while we exclusively focus on bug fixing agentic-PRs. Combining all types of changes might hide relevant reasons specific to fixing issues and under-represent specific reasons that are more relevant for fixing bugs and less relevant for other types of task. Focusing on only bugs will guide future studies on providing customized solutions for fixing bugs, which is aligned with the whole software engineering literature that has distinct sub-fields for fixing bugs, creating new features, writing high quality documentation, etc.. For example, a previous work~\cite{pantiuchina2021developers} specifically looked at the rejection of pull requests for refactoring. (3) The prior work on Claude manually studied only 92 rejected PRs, while we studied a larger and representative sample of 306 rejected fixes. Replicating the findings on a larger dataset on itself without the other differences is necessary for PR acceptance~\cite{chen2019replication}.

In this paper, we perform a qualitative analysis on a representative sample of 306 rejected AI-agentic fixes to identify the reasons for the rejection and a quantitative analysis to better understand these reasons. 
We addresses the following research questions: 
    
\noindent \textbf{RQ1. \rqi}
We qualitatively investigate the reasons for rejecting AI-agent fixes. We find that rejected issues are due to incorrect requirements, incorrect implementation (e.g., wrong approach), technical issues related to the fix (e.g., failing CI), or the fix itself is low priority.

\noindent \textbf{RQ2. \rqii}
We quantitatively investigate the wasted effort in terms of code churn and exchanged comments, associated with the rejection of AI-agent fixes. We find that those fixes 
incur a substantial amount of code churn ranging from a median of 81 to 293.

\textbf{Take-home message:} Our findings suggest that developers should better 
guide the agents on how to approach the issue, identifying the approaches that can be considered invalid, discussing how to validate the fixes so that they pass tests and do not break CI pipelines, and minimizing the use of AI-agents for low priority fixes. Our findings suggest further studies on how to design and phrase prompts for fixing issues.

Our replication package is made publicly available~\cite{replication-package}.

\section{Data Collection}

The study uses the AIDev dataset \cite{li2025aiteammates_se3}, which includes pull requests generated by AI coding agents across various open-source GitHub repositories. We leverage the AIDev dataset with more than 100 stars, comprising 33,596 curated pull requests distributed over 2,807 projects. As we wish to better understand the reasons for rejecting agent-generated fixes, we consider pull requests that are (i) closed and not merged, (ii) created or co-authored by Copilot, Devin, Cursor, or Claude, and (iii) propose agent code fixes. We exclude Codex pull requests since they are just labeled with ``codex'', and we cannot systematically identify their contribution to the fixes. This results in 1,497 rejected AI-agentic fixes, from which we select a representative random sample (confidence level = 95\% and margin of error = 5\%) of 306 pull requests to study. The AIDev dataset features a total of 3,225 fix pull requests contributed by the four agents, out of which 1,497 were rejected fixes, forming \textbf{46.41\%} of the fixes.

\section{Results}

\textbf{RQ1. \rqi}

\begin{table*}[t!]
\small 
\centering
\footnotesize
\caption{Taxonomy of rejection reasons for AI-Agent pull requests.}
\label{tab:rejection-reasons}
\begin{tabularx}{\textwidth}{l l X X c} 
\toprule
\textbf{Category} & \textbf{Sub-category} & \textbf{Definition} & \textbf{Example} & \textbf{\# (\%)} \\ 
\midrule

\multirow{10.5}{*}{\parbox{2.2cm}{Relevance of the fix}} 
& Inactivity & The pull request was automatically closed after a predefined period triggered by the absence of interaction. & \textit{``Closing due to inactivity for more than 7 days''} & 53 (17.3\%) \\
\cmidrule{2-5}
& Superseded & PRs that were closed because they were replaced by other PRs or a better alternative. & \textit{``Closed in favor of \#1356''} & 18 (5.9\%) \\
\cmidrule{2-5}
& Low priority & The PR was closed because the issue is minor or non-critical. & \textit{``This seems like a low prio issue [...]''} & 3 (1.0\%) \\
\cmidrule{2-5}
& Architecture & Closed because the architecture or framework changed significantly. & \textit{``[...] major architectural change [...]''} & 1 (0.3\%) \\
\cmidrule{2-5}
& Test PR & PRs that were created only for testing purposes. & \textit{``test PR''} & 1 (0.3\%) \\
\cmidrule{1-5} 

\multirow{11}{*}{\centering\parbox{2.2cm}{Implementation issues}} 
& Incorrect fix & PRs closed due to functionally incorrect or incomplete changes. & \textit{``this PR is not doing what it says [...]''} & 17 (5.6\%) \\
\cmidrule{2-5}
& Wrong approach & PRs closed because the issue was resolved through an alternative implementation. & \textit{``@copilot -- went a different way [...]''} & 8 (2.6\%) \\
\cmidrule{2-5}
& Ambiguity & Changes based on an incorrect understanding of requirements. & \textit{``I think it's confused because [...]''} & 2 (0.7\%) \\
\cmidrule{2-5}
& Insufficient & The PR partially fixes the issue but leaves key functionality incomplete. & \textit{``[...] the detail property is not provided''} & 2 (0.7\%) \\
\cmidrule{2-5}
& Wrong repo & Submitted to the wrong repository. & \textit{``Plz PR to eliza-plugins org [...]''} & 2 (0.7\%) \\
\cmidrule{1-5}

\multirow{3.5}{*}{\parbox{2.2cm}{Provider}} 
& Agent failure & AI agent failed, became unreachable, or produced erroneous output. & \textit{``Devin is currently unreachable [...]''} & 23 (7.5\%) \\
\cmidrule{2-5}
& Rate Limit & Could not be executed due to API or resource restrictions. & \textit{``[...] hit a rate limit [...]''} & 3 (1.0\%) \\
\cmidrule{1-5}

\multirow{3.5}{*}{\parbox{2.2cm}{Technical issues}} 
& CI failure & PRs closed because CI pipelines or automated tests failed. & \textit{``build still fails [...]''} & 21 (6.9\%) \\
\cmidrule{2-5}
& Breaking change& Introduced unintended side effects or broke existing functionality. & \textit{``This implementation introduces a breaking change.''} & 1 (0.3\%) \\
\cmidrule{1-5}

Others & Others & N/A & N/A & 151 (49.3\%) \\
\bottomrule
\end{tabularx}
\end{table*}

\textbf{Motivation:} This research question aims to deepen our understanding of the reasons developers reject agent-suggested code fixes. Such insights can help practitioners leverage AI-based coding agents more effectively, clarifying when such agents fall short and guiding future research on how to improve the integration of AI-agents as reliable teammates. 

\textbf{Approach:} We adopt a qualitative analysis technique to identify the main reasons for rejecting AI-agent code fixes. Two authors independently label the 306 pull requests, by 
carefully reading the pull request discussion threads (i.e., commensts, CI results, external artifacts) to extract evidence explaining the rejection of AI-agent fixes. 
Second, based on the derived context, each author assigns an initial label that captures the underlying reason for rejection. Through a discussion, we 
come up with a final set of labels, which is used to label all the data. With the final labels, 
we assess the inter-rater reliability between the two raters using Cohen's $\kappa$, achieving a value of 0.605, which indicates substantial agreement. To resolve the disagreement cases, the third author conducts an analysis on the 93 conflicting cases reviewing them to reach a consensus. Then, all the three authors together through a discussion regroup the obtained labels into high level ones. 

\textbf{Results:} We identify four high-level themes underlying the rejection of AI-agent fixes, which are further decomposed into 14 sub-themes. Table~\ref{tab:rejection-reasons} summarizes these themes along with their definitions and illustrative examples, and we discuss them in detail in the following findings. Note that we were unable to label 151 agentic-PRs since they have no clear reason for rejection.

\textbf{Relevance-related issues are the most common reason for rejecting AI-agent fixes.}
This category captures agentic pull requests whose proposed fixes become less relevant over time. This high-level theme is further divided into five sub-themes. The most prevalent sub-theme belongs to pull requests that remain inactive for long periods~\cite{instructor_pr1554}, accounting for 17.3\% of the cases. We also observe that superseded pull requests~\cite{autoscaling_pr1353} form the second emeging sub-theme with 5.9\%, typically occurring when other pull requests address the same issue, propose a better alternative, or replace the original fix. The remaining sub-themes are less frequent and include low-priority pull requests (three cases)~\cite{curator-pr305}, one case related to architectural changes~\cite{claude-flow-pr149}, and one for testing purposes~\cite{azure-sentinel-pr12466}. 

These results highlight the need for an approach that prioritizes issues to fix in order to avoid creating duplicates or pull requests that remain inactive with no further activity. These can be seen as waste in terms of agentic-AI resource consumptions (e.g., compute, tokens for token-based models and premium queries for agents like Copilot). They are also similar to the abandonned changes that are well known in the literature as an important waste~\cite{WANG2019108, 10.1145/3530785, 9332267}.

\textbf{AI-agent fixes can be rejected when they are not properly implemented, forming the second major rejection theme.} As shown in Table~\ref{tab:rejection-reasons}, implementation-related issues refer to changes that follow an incorrect approach and are divided into five sub-themes. The most common sub-themes involve pull requests closed due to an incorrect fix~\cite{maui-toolkit-pr219} or an inappropriate solution approach~\cite{rqlite-sql-pr75}, accounting for 5.6\% and 2.6\% of the qualitatively analyzed cases, respectively. Similarly, AI-agent fixes are rejected when requirements are ambiguous~\cite{typescript-pr61902} or not adequately addressed~\cite{turborepo-pr10408}, or when changes are made in the wrong repository~\cite{owncast-pr4354}.

These findings suggest that clearer guidance on preferred solutions—and explicit constraints on undesirable approaches—can better guide AI agents. This guidance can be provided through project-specific instruction files~\cite{githubdocs_copilot}, or explicitly included in prompts or GitHub issues related to the PRs.

\textbf{Technical-related issues constitute another category of AI-agent pull request rejections,} comprising two sub-themes: \textit{CI failures} and \textit{breaking changes}, accounting for 21 and 1 PRs, respectively. 
In CI-related rejections, we observe cases where tests fail or infrastructure issues prevent the agent from successfully landing the fix. A good example is shown in Table~\ref{tab:rejection-reasons}, where the CI pipeline repeatedly failed during the build process~\cite{apiclientcodegen_pr1199}. In contrast, only one case involved an implementation that introduced a breaking change~\cite{azure_sdk_net_pr50357}. 

Consequently, this category indicates that AI agents often lack sufficient contextual information on how to write, configure, and execute tests that pass successfully. Providing agents with clearer testing guidelines and CI-related instructions may help reduce such rejections and improve task execution.

\textbf{Provider-related issues are also a common reason for rejecting AI-agent pull requests.} 
Agents may fail to complete tasks due to provider-side problems, such as becoming unreachable, producing incorrect responses, or failing during execution; this sub-theme accounts for 23 pull requests. Note that we also consider pull requests with no commits in this category. For example, the Devin agent encountered issues when responding to a developer's query~\cite{crewai_pr3113}, as reflected in the message: \textit{``Devin is currently unreachable—the session may have died. Let's just do that and then close this one.''}. Second, provider-imposed rate limits can prevent agents from continuing their work, a sub-theme observed in three cases. For instance, Copilot reached the maximum number of allowed requests~\cite{taigaui_pr11352}, as noted in the message: \textit{``Sorry, you've hit a rate limit that restricts the number of Copilot model requests [...].''}. About the rate limit of Copilot, any request to the agent can be seen as a new request, which quickly sums up and reach the limit. These findings suggest that provider-specific constraints can directly impact the success of AI-agent pull requests.

\summarybox{Summary}{
We observe 14 reasons for rejecting Agentic-PRs for fixing bugs. They are related to the implementation that can be incorrect, 
face issues on the CI pipeline, the agent failed to perform the task, or the fix is 
rejected for irrelevance.
}

\textbf{RQ2. \rqii}

\textbf{Motivation:} The goal of this research question is to quantify the number of changes made by agents that end up being rejected for each of our five identified reasons from RQ1. The greater the number of changes, the more effort is required to review the change, and the higher AI-agents resources are wasted. 

\textbf{Approach:} To answer this RQ, we collect the code churn and the number of commits for each pull request using the AI-Dev dataset and compare the five categories with each other.

\begin{figure}[!h]
\centering
\begin{subfigure}[b]{0.23\textwidth}
    \centering
    \includegraphics[width=\textwidth]{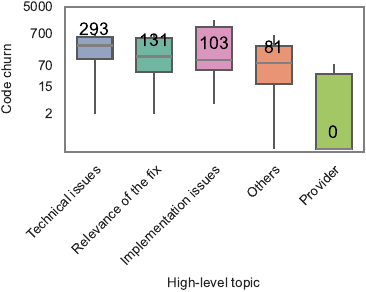}
    \caption{Code churn}\label{fig:code-churn}
\end{subfigure}
\hfill
\begin{subfigure}[b]{0.23\textwidth}
    \centering
    \includegraphics[width=\textwidth]{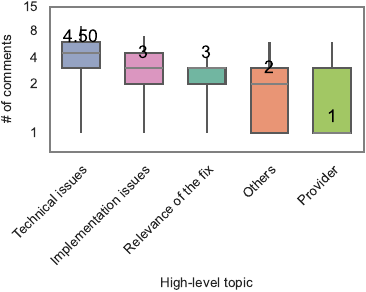}
    \caption{Comments}\label{fig:comments}
\end{subfigure}
\caption{The code churn and discussion comments across rejected PR topics.}
\end{figure}

\textbf{Results:} \textbf{Rejected PRs have a large code churn with a median that ranges from 81 to 293}, as shown in Figure~\ref{fig:code-churn}. The Provider category has a median of zero, since we consider changes with zero commits in that category (the agent was unable to generate code). We observe that the category with the highest code churn is related to the pull requests that face technical issues. That can be explained by the fact that the larger the change, the more likely it will lead to build failures. That can suggest further studies to look into these failing tests. One assumption is that the AI-agents with a large amount of changes might need to adjust the tests as well. The agents should also have guidance on how to handle these failures, by potentially providing tests that should be run for a given fix. The implementation issue can also introduce an important waste as developers need to review a large amount of changes (a median of 103 lines) to find out that the generated fix is inadequate.


\textbf{Although we do not observe a large number of comments, the rejected pull requests still received comments}. In fact, the median number of comments is between one and 4.5, as depicted in Figure~\ref{fig:comments}. However, this number is not different from that of the merged fixes. The merged fixes have a median of two comments (Q1 and Q3 of one and 4). As such, and since the number of rejected pull requests constitutes almost half of the data (46.41\% as discussed in the Data Collection Section), almost half of the comments are wasted. Even for tasks that are irrelevant, they have a median of three comments before rejecting the pull request. 

\summarybox{Summary}{
Rejected Agentic-PRs have a median code churn that range between 81 and 293. The higher code churn is associated with the technical issues (e.g., CI-failure), followed by the Agentic-PRs rejected for their relevance then implementation issues. 
Rejected fixes involve a median of one to 4.50 comments similarly to the merged ones (a median of of two).
}

\section{Implications}

This section discusses some implications of our findings.

\textbf{Implication 1: We suggest that developers provide guidance on how to perform the fix}. We observe that a good amount of rejected fixes are due to an incorrect way of fixing issues. As such, we suggest that developers provide guidance on how to perform the fix or provide guidance on what approaches are not acceptable in the agent instruction file (e.g., .github/copilot-instructions.md).

\textbf{Implication 2: We suggest developers provide agents with instructions on how to validate a change.} A good amount of changes are rejected because the issue is not fixed, fixes fail tests or CI pipelines. Guiding the agent on how to run tests and how to validate if the issue is fixed or would not avoid agents generate inadequate fixes. That is in addition to further studies on understanding the CI failures and how to instruct agents to avoid them.

\textbf{Implication 3: Better task management is required to minimize using AI-agents for low priority tasks that end up being closed}. These tasks despite their low priority, agents generate a large amount of changes for them which require human reviews and testing such that they do not break the code. Irrelevant and low priority fixes should be avoided at the first place to avoid generating irrelevant code that would also involve human reviews. Although a large body of prior research \cite{10.1145/3131704.3131725, 10.1145/3196398.3196455} focused on duplicate patches and pull requests, these studies need to be replicated to avoid agents creating duplicate fixes from the prompt itself. As such, studies are required to evaluate the priority of tasks all the way from their requirements before assigning them to AI-agents.

\textbf{Implication 4: We suggest researchers develop approaches (e.g., templates) that better guide developers prompt the AI-agents for fixing issues;} Our study enumerates a set of reasons for rejecting the AI-agents fixes, which can motivate future studies to develop quality metrics that assess the quality of a prompt such that the AI-agent implements a successful fix. For example, does the issue enumerate its relevant CI tests? 
\section{Conclusion and Future Work}

46.41\% of the fixes created by Copilot, Devin, Cursor, and Claude are rejected. Understanding the rejection of these fixes is critical to determine AI-agents fall short, such that developers can better integrate agents in their pipelines and researchers can provide solutions to guide developers on better use of agents to fix issues autonomously (aka., \textbf{autonomous AI-agents}). 

Toward this goal, we conduct a qualitative study on the reasons for rejecting fixes. We find four high-level reasons that encourage developers to provide guidance on how to perform the fix, guidance on how to validate if the change fixed the issue, and guidance on how to prioritize relevant issues (not low priority ones). The categories obtained are likely to involve a good amount of wasted resources, coming from the amount of generated code (AI-agent resources) and human efforts to review the generated fixes. Almost half of the comments are around rejected fixes. 

In the future, we plan to design approaches that can better guide developers in using AI-agents for successfully fixing issues. For instance, we plan to provide and evaluate a prompt template, evaluate the contribution of instruction files on improving the AI-agents, design approaches that recommend tests relevant to an issue, and that can be used as a validation point for generated fixes.


\bibliographystyle{ACM-Reference-Format}
\bibliography{sample-base}

\end{document}